%% file: main-sigconf.tex
\documentclass[sigconf]{acmart}

\usepackage{amsmath,amsfonts}
\usepackage{algorithmic}
\usepackage{graphicx}
\usepackage{textcomp}
\usepackage{xcolor}
\usepackage{listings}
\lstdefinestyle{customasm}{
  belowcaptionskip=0pt,
  captionpos=b,
  breaklines=true,
  frame=single,
  xleftmargin=7pt,
  xrightmargin=3pt,
  numbers=left,
  language=[x86masm]Assembler,
  commentstyle=\itshape\color{purple!40!black},
  morekeywords={LDADR, BRLET, COD, MEMCOD, JMPPC, STADR, sum, BREQ, LDOFF, STOFF, JMPLBL},
  morekeywords=[2]{Addr, COMMIT, reg, vecAdd_2048L},
  morekeywords=[3]{def, enddef},
  morekeywords=[4]{R8b_1, R8b_2, R8b_3, R64B_1, R64B_2, R64B_3, R64B_4, R64B_5, R64B_6, R64B_7, R64B_22, R64B_11, R2048L_1, R2048L_2, R2048L_3},
  keywordstyle={\color{blue}\textbf},
  keywordstyle=[2]{\color{orange}\textbf},
  keywordstyle=[3]{\color{blue}\textbf},
  keywordstyle=[4]{\color{teal}\textbf},
  basicstyle=\footnotesize\ttfamily,
  numberstyle=\color{blue}\textbf
}
\usepackage{subcaption}
\usepackage[font=small,labelfont=bf]{caption}
\usepackage{setspace}
\AtBeginDocument{%
  \providecommand\BibTeX{{%
    \normalfont B\kern-0.5em{\scshape i\kern-0.25em b}\kern-0.8em\TeX}}}

\setcopyright{acmcopyright}
\copyrightyear{2018}
\acmYear{2018}
\acmDOI{XXXXXXX.XXXXXXX}

\acmConference[PPoPP - ExHET 2023]{Make sure to enter the correct
  conference title from your rights confirmation emai}{February 25 -- March 01,
  2023}{Montreal, Canada}
\acmPrice{15.00}
\acmISBN{978-1-4503-XXXX-X/18/06}

\begin{document}

\title{On Memory Codelets: Prefetching, Recoding, Moving and Streaming Data}

\author{Dawson Fox}
\email{dfox@anl.gov / dawsfox@udel.edu}
\affiliation{%
  \institution{Argonne National Laboratory}
  \city{Lemont}
  \state{Illinois}
  \country{USA}
}
\affiliation{
    \institution{University of Delaware}
    \city{Newark}
    \state{Delaware}
    \country{USA}
}

\author{Jose Monsalve Diaz}
\email{jmonsalvediaz@anl.gov}
\affiliation{%
  \institution{Argonne National Laboratory}
  \city{Lemont}
  \state{Illinois}
  \country{USA}
}

\author{Xiaoming Li}
\email{xli@udel.edu}
\affiliation{%
  \institution{University of Delaware}
  \city{Newark}
  \state{Delaware}
  \country{USA}
}


\begin{abstract}
For decades, memory capabilities have scaled up much slower than compute capabilities,
leaving memory utilization as a major bottleneck.
Prefetching and cache hierarchies mitigate this in applications with easily predictable
memory accesses or those with high locality. In other applications like sparse linear algebra
or graph-based applications, these strategies do not achieve effective utilization of memory.
This is the case for the von Neumann model of computation, but other program execution models (PXM) provide different
opportunities. Furthermore, the problem is complicated by increasing levels of heterogeneity
and devices' varying memory subsystems. The Codelet PXM presented in this paper provides a program structure that allows
for well-defined prefetching, streaming, and recoding operations to improve memory utilization
and efficiently coordinate data movement with respect to computation. We propose the Memory Codelet, an extension to the original Codelet Model,
to provide users these functionalities in a well-defined manner within the Codelet PXM.
\end{abstract}



\keywords{Codelets, Program Execution Models, Sequential Codelet Model, Heterogeneity, Memory Recode, Near Memory Compute}


\received{22 December 2022}

\maketitle

\input{tex/intro}
\input{tex/codeletModel}
\input{tex/memCodelets}

\input{tex/conclusion}



\bibliographystyle{ACM-Reference-Format}
\bibliography{bib/capsl, bib/CodeletModelforChiplets, bib/MemoryCodelets}



\end{document}

%% file: tex/intro.tex
\section{Introduction}

Compute capabilities on a single chip increased rapidly for decades as a
result of Moore's Law and Dennard Scaling. Memory performance, however, is increasingly becoming the first order constraint. There are two main causes behind the so-called ``memory wall''. Firstly, memory latency and bandwidth have increased much slower than their logic circuit counterparts \cite{SRAMScalingIsDead}. Secondly, cache hierarchies and prefetching mechanisms, two of the industry's
most prevalent answers to this problem, naturally lost their benefits when the memory accesses became more concurrent and less predictable.
As a result, modern day applications with less common memory access patterns are often memory bound, and the memory wall remains a major obstacle on the systems of today. 
Graph-based and sparse linear algebra applications, particularly their parallel implementations, are examples 
that can have poor performance on conventional systems due to concurrent pointer chasing and irregular memory accesses.

Recently, computer architects have been exploring innovative ways of overcoming the 
memory wall. Smart prefetching \cite{ptask,prodigy,ImprovedDMCacheWPrefetchBuffers,IndirectMemoryPrefetcher,Minnow2018}, near-memory and in-memory 
computing \cite{tensorDIMM,RecNMP,ScalablePIMforParallelGraphs}, recoding engines \cite{UAP,UDP2017,tako,ProgrammableAccelerationforSparseMats},
and other techniques have been proposed to accelerate applications past
the memory wall. An approach similar to ours has been attempted \cite{EfficientDataSupplyParallelHetArchs} but not involved with or benefiting from
the Codelet Model PXM,
building off of \cite{DAEComputerArchs} instead.
As the architectures become more heterogeneous, the problem of orchestration 
of computation and data becomes more challenging. Furthermore, memory-focused accelerators bring additional
problems that would be difficult to tackle with current execution models. However, the orchestration of 
computation across a highly heterogeneous system that features these innovations is typically left to the
software developer. The status-quo software-handmaid solution is simply not productive nor scalable. Our key insight is that a Program Execution Model (PXM)
\cite{ParallelTuringMachine, ModularSoftwareConstructionDennis} for 
extremely heterogeneous systems, supporting both compute and memory semantics, will greatly relieve the developer's burden and likely lead to higher productivity and efficiency.
This paper presents the concept of Memory Codelets, a definition that extends from previous versions
of the Codelet Model \cite{TM104,monsalveExHET,MonsalveIDPRM2019,SuetterleinZucGao13} with a focus on 
smarter memory orchestration in the presence of extreme heterogeneous architectures.
\vspace{-1em}
\subsection{Motivation}
Currently, the most common orchestration method for accelerators is the offloading model,
similar to the organization of execution commonly seen in heterogeneous systems that
make use of GPUs and other accelerators. 
This model distinguishes a host, usually the CPU where an operating system runs, and multiple 
devices called accelerators. Work is divided up into kernels
that the host deploys into accelerators. Data is transferred to/from the accelerator's 
memory and the host memory. While most accelerators provide manual memory management capabilities, 
some modern systems are also implementing unified shared memory to maintain automatic 
coherence between accelerator and host memories.

Orchestration of programs in the offloading model is a manual process. The user must coordinate 
scheduling of data and kernels into the different accelerators. From the perspective of the host,
a kernel is just a device function initiated through the driver API, and the user or system
must guarantee that data needed for its execution is appropriately allocated and
initialized prior to the execution of the kernel. A kernel itself does not provide any information that explicitly 
mentions the relationship about the memory it requires to perform its computation. Therefore, the user relies on 
manual memory management and system responses to page faults or cache faults to coordinate memory needed for the execution of a kernel.
As a consequence, the scheduler does not have any knowledge of memory operations, or how a kernel depends
on specific data, instead, memory and compute must be organized in a meticulous way that respects such dependencies. As data movement grows more important for achieving high performance, the burden on the
programmer grows greater.

Existing alternatives to the offloading model allow kernels/tasks to have unrestricted behavior and lack
clear definitions of the data consumed and produced, which does not aid in predicting memory accesses
or utilizing the memory systems better. 
The common alternative is heterogeneous tasking models \cite{DesignOpenMPTasks2009, TheOngoingEvolutionOpenMP22018, OmpssCluster2022, TaskMappingHeterogenousCPUGPU2021, OmpSsPTaskBasedApplications2014, cudaGraphsGettingStarted}. 
Kernels are represented as tasks that are connected by dependencies forming a graph. However, tasks are not side-effect free, as they are part 
of a unified shared memory system, allowing any pointer within the task to reference any part of the memory. Furthermore, 
models often use dependencies only to represent control flow dependencies, not data dependencies. Thus, although 
memory locations (e.g. variables or pointers) are used to name dependencies, these are not necessarily used to 
make decisions about memory orchestration, nor do they represent all the set of memory locations that can be accessed by the task. Take for example an OpenMP \texttt{target} region that uses the 
\texttt{depend} clause to define dependencies, and the \texttt{map} clause to define data movement. The depend clause only determines
the producer-consumer relationship of tasks, while the map clause only determines memory 
movement operations between host and device. However, a task may contain pointers not defined in the \texttt{depend} clause, for example when dealing with global variables. This is worse when a unified shared memory system spans across host and device, since pointers in the tasks
may interact with any part of the subsystem \cite{AutomaticAsyncOffloading2022LLVMHPC}. Such freedom makes it hard to predict task latency and side effects. As a result, performing operations such as streaming, data recoding, or using heterogeneous systems with memory accelerators, are not well differentiated in the program description and execution, and must be managed by the user. Saying it differently, the scheduler is not in charge of
orchestrating memory and its relationship to compute. Restricting tasks to always define dependencies helps. Such an approach is used in the OpenMP Cluster model \cite{ClusterProgrammingOpenMP2018,TheOpenMPClusterProgramming2022}, allowing the runtime to perform smart memory management. However, this does not yet resolve issues with complex memory hierarchies found in heterogeneous systems.

Previous work has demonstrated that the Codelet Model can be
an effective programming model for heterogeneous systems \cite{TongshengPDAWL2020, CodeletModelinChiplets, SuperCodeletArchExHET}.
However, Codelets being non-preemptive can be both a strength and a weakness. Without the
ability to interrupt Codelets, data locality becomes
essential when designing a high performance and efficient program. If required data is
not local to a Compute Unit (CU) when a Codelet is executed, the processor effectively stalls while
the data is fetched. With proper management, on the other hand, Codelets' atomic nature can
ensure that computation is performed while the necessary data is local and that the
data need only be local for as short a time as possible. This missing capability is the key to realize the Codelet model's full potential on extreme scale heterogeneous systems.
Prior work has suggested percolation as an important way to improve performance
through the memory wall \cite{tan2008percolation}.
Our proposal of the Memory Codelet provides exactly such an 
explicit mechanism to orchestrate memory for compute Codelets. Memory Codelets will be executed on
a near-memory architecture dedicated to Memory Codelets only, and would be able to prefetch data,
stream data, and perform recode operations. More specific operations like pointer swizzling might also
be handled by Memory Codelets. On systems with less conventional memory hierarchies or
with multiple devices, Memory Codelets would also be able to explicitly move data throughout the
memory hierarchy (for example, to local scratchpad memory) and coordinate reads, writes, streams,
and recode operations to the various devices.


\subsection{Contributions}

The contributions of this paper are described as follows:
\begin{itemize}
    \item Conceptualization of Memory Codelets and their integration into the Codelet Model
    \item Integration of Memory Codelets into the Codelet Model, the Sequential Codelet Model, and their Abstract Machines
    \item Simple examples that demonstrate the use of Memory Codelets through Sequential Codelet Model semantics
\end{itemize}

%% file: tex/codeletModel.tex
\section{Background}
Program Execution Models (PXMs) as a concept are effectively a holistic system view
that offers
clear and well-defined behavior at all levels of the system from the hardware to the
software. As this is often difficult to achieve without significant resources
and end-to-end design, PXMs
are often relegated to the domain of software runtimes in practice; however, they still
offer the user/developer an organization of execution that can be relied upon and used
to build effective programs for systems. This is especially important as the industry
trends towards extreme heterogeneity, and various programming models and APIs might be used
to craft a single high performing program. A well defined program execution model is a fundamental
step towards hardware/software co-design \cite{ang2022reimaginingCodesignDoE}. A more precise definition of PXMs can be found
in \cite{ParallelTuringMachine,CAPSL2011,Dennis97}.

\subsection{Codelet Model}
The Codelet Model is a dataflow-inspired PXM to organize computation with an
accompanying abstract machine \cite{TM104,SuetterleinZucGao13}.  It is both fine-grained and event driven, breaking
programs into Codelets, non-preemptive portions of 
sequential computation with defined inputs and outputs. As
such, Codelets are the quantum unit of scheduling of a Codelet-based program. Programs in the 
Codelet Model are described by Direct Acyclic Graphs, with nodes in the graph representing Codelets
and directed arcs representing data and control dependencies between them. This allows the program to clearly define the
ordering of Codelet execution only where necessary while permitting flexibility in Codelet scheduling
otherwise. Codelets are also event driven, allowing dependencies to be seen as split phase transactions, or 
as activations based on external events. 
To benefit more from locality, Codelets are grouped into Threaded Procedures (TPs). The
Codelet Abstract Machine defines the components of a system that executes Codelet Programs.
It designates the roles of Compute Unit (CU) and Scheduling Unit (SU):
the SU is responsible for creation of TPs and scheduling Codelets during runtime, while the CU
is responsible for "firing" (executing) Codelets once their dependencies are fulfilled. 
For a more fleshed out description of the basic
Codelet PXM, view prior publications \cite{DARTSEuropar, CodeletModelinChiplets}.

Since CUs are general in the abstract machine, almost any computational architecture can be mapped to them,
which enables the Codelet Model PXM to gracefully handle extreme heterogeneity. As long as Codelets contain
computation that can feasibly be performed on at least one CU of the system or has a version that can be
executed on that CU, heterogeneous scheduling in the Codelet Model implementation can enable dynamic heterogeneity
as in \cite{TongshengPDAWL2020}. The event-driven dependency-based scheduling can aid users to reason about synchronization
between the Codelets of the program, whereas typical heterogeneous execution on conventional systems leaves 
synchronization and explicit data movement to the user. Such a burden can slow the development process and lead to
hours of debugging. Furthermore, with systems increasingly having various specialized architectures, computation of
individual tasks (or in this case, Codelets) may be drastically sped up. In some applications this may lead to 
increased strain on the memory architecture, which indicates that better memory management and data movement
throughout the memory hierarchy is paramount to continue improving performance.

\subsection{Sequential Codelet Model}

The Sequential Codelet Model (SCM)
\cite{monsalveExHET,MonsalveIDPRM2019,MonsalvePhDThesis} is an extension of the
original Codelet Model \cite{TM104,SuetterleinZucGao13}. SCM is heavily inspired by
instruction level parallelism (ILP) techniques in sequential computing architectures.
In ILP, dataflow is used to discover dependencies in the instruction stream and allow
parallel execution of independent instructions. In particular, SCM heavily draws from
out-of-order execution that uses register names to respect true dependencies while removing
anti- and output dependencies.

In SCM, Codelet graphs are defined as a sequential stream of instructions where
dependencies are registers. However, compared to traditional registers in sequential
architectures, the dependencies between Codelets are larger in size. This is possible
because the pipeline in charge of executing the Codelet program (i.e. Fetch, decode,
execute, memory and write back) sits on top of the compute units.
Therefore, registers are assigned to a location equivalent to
the last level of cache (LLC). The execution stage of the
pipeline is comprised of these CUs, which can be implemented as any architecture such as
Streaming Multiprocessors (in GPGPUs),
a single or multi core CPU system, an FPGA, or any other exotic/specialized architecture. 

Control flow instructions are also supported in SCM. These instructions allow 
dynamically defining more complex Codelet Graphs by using conditional and unconditional
jumps, loops, and a subset of basic arithmetic operations. Additionally, memory
operations are necessary to move data back and forth between the upper level memory
(e.g. DRAM) and the register file. In this work we formalize this concept as Memory
Codelets, which is equivalent to memory operations in sequential ISAs, yet more
powerful thanks to the possibility of Memory Codelets to be user defined. A complete description of the Sequential Codelet Model and its realization in the SuperCodelet architecture can be found in \cite{MonsalvePhDThesis, monsalveExHET, MonsalveIDPRM2019}.

%% file: tex/memCodelets.tex
\section{Memory Codelets and the Codelet Abstract Machine}

Traditional instruction set architectures (ISA) group instructions according to their functionality. 
Such distinction also has an effect in the functional unit used in the pipeline of the architecture 
that implements the ISA. Among the different groups, memory instructions (e.g. load, and store 
operations) focus on the interaction and movement of data in and out of the compute pipeline. In addition
to ISA instructions, current system architectures feature caching and prefetching mechanisms with the purpose 
of managing memory in the system, such that it increases performance while respecting a set of rules enforced
by memory consistency and coherency models. Prefetching, for example, is an event driven mechanism that is triggered 
based on multiple access patterns across multiple memory requests. In addition to the ISA, conventional systems
typically have multilevel cache hierarchies that are completely invisible from the
software perspective and do not provide reconfigurability nor programmability in
the memory hierarchy. The rigidity of the coordination of memory in the hierarchy does not allow enough
flexibility to execute programs with uncommon memory access patterns at high
performance.

In comparison, the organization
of programs in the Codelet Model gives a definite and specific plan of what data is
consumed/produced by each Codelet, which provides the information necessary for
prefetching \cite{prodigy}, recoding \cite{UDP2017,tako}, and streaming strategies 
that are well integrated into the abstract machine and provides customization of the memory 
management. By virtue of the Codelet
Model PXM, these strategies would benefit most from being implemented at multiple levels
of the system stack and employing hardware/software co-design. More benefits of the
Codelet Model PXM in future extremely heterogeneous architectures are summarized in \cite{CodeletModelinChiplets}.

Memory Codelets in the Codelet Abstract Machine have a similar role for memory management as the
prefetching mechanism does in conventional execution. 
Memory Codelets allow for data to be moved, manipulated, reorganized or streamed through different
components in the system, aiming to serve compute Codelets such that their execution time is driven 
by arithmetic operations rather than memory accesses. Memory Codelets make memory management explicit, 
and they are particularly useful when memory access patterns cannot be recognized in the memory subsystem. 
This is the case for applications that cannot easily take advantage of caches and prefetching. Like 
traditional Codelets, Memory Codelets are event- and data-driven, and form independent nodes in the Codelet graph. However, instead of mapping to compute units that are heavily specialized for arithmetic operations, memory Codelets map to specialized units that emphasize throughput and low latency \cite{UDP2017,UAP,tako,UpDownAgileArticle}.

\subsection{Memory Codelet Abstract Machine}
\begin{figure}[h]
    \centering
    \includegraphics[width=8cm]{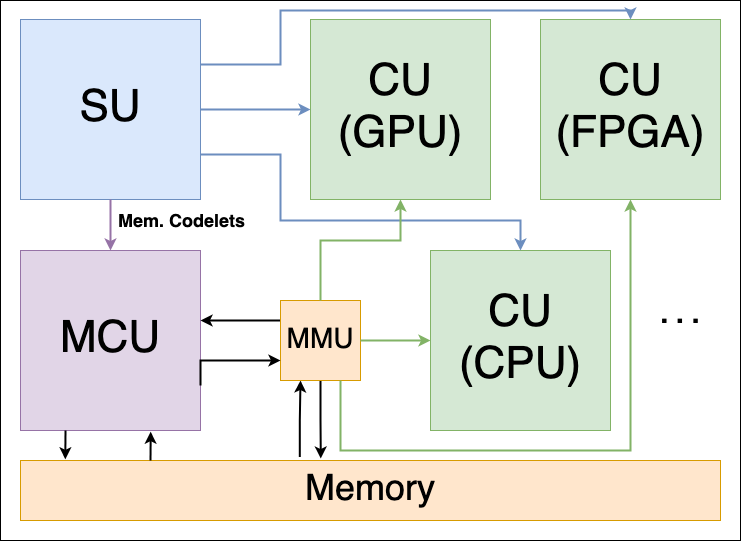}
    \caption{Memory Codelet Abstract Machine with possible heterogeneous CU architectures
    mentioned.}
    \label{fig:memory_codelet_CAM}
\end{figure}

Let us begin with a definition of an abstract machine that extends from the Sequential Codelet Model, and the
original Codelet Model. \autoref{fig:memory_codelet_CAM} shows the different components of the Memory Codelet 
Abstract Machine. We focus on a single level of machine hierarchy, while this can be extended to other levels
of the machine abstraction. As in traditional Codelet Models, the architecture is made out of Compute Units 
(CUs) and Scheduling Units (SUs). Additionally, a Memory Codelet Unit (MCU) is included, that is in charge of execution
of memory codelets. There are two aspects that make the Memory Codelet Unit different to any other compute unit.
First, its compute capabilities are tailored for fast data transformation (e.g. \cite{UDP2017,tako,RecodeWebsite}) and movement.
Second, it can directly interact and communicate with the different memory storage components of the system, including local 
memory, external memory, and specialized memory structures (e.g. FIFO queues). 

Because the Codelet Model clearly defines Codelets' inputs and outputs, Memory Codelets
can be leveraged to benefit from the static Codelet graph of the program being
executed. The Codelet Graph defines a partial execution order of the Codelets based on
their data dependencies.  The actual execution order of
Codelets depends heavily on the scheduling mechanism employed in the Codelet Model
implementation; for example, DARTS \cite{DARTSEuropar}, a software runtime
implementation of the Codelet Model, employs multiple scheduling mechanisms such as round robin and
work stealing that the developer can choose for their program. These scheduling 
mechanisms always respect data dependencies to ensure correct program execution.

\subsection{Prefetching, Streaming, and Recoding with Memory Codelets}

When a Memory Codelet is executed, data does not need 
to always pass through the MCU silicon fabric. The data can be moved directly between different physical locations
in the system (e.g. CU to CU or DRAM to a CU directly). Thus, the MCU does not have to be a bottle neck for memory transactions,
but a coordinator for memory across the Codelet Abstract Machine. Since Memory Compute Units can be seen as
traditional near memory compute architectures, data can be fetched to the MCU, transformed, and delivered to other
physical locations. In keeping with the goals of a PXM, the MCU is designed to understand Codelet semantics
for scheduling and dependencies management.

This mechanism can be used
to ensure data locality prior to the scheduling and execution compute Codelets. This can be an effective prefetching
mechanism, especially given that memory Codelets are written by the developer who has
knowledge of the access patterns of the program. Thus, Memory Codelets can perform the necessary
preprocessing to determine the data needed for the compute Codelet, even if these operations are complex. For example, data does not need to be contiguous or respect a simple stride pattern. A Memory Codelet can perform pointer chasing across the graph to obtain the necessary properties from different nodes.
With the Memory
Codelet Unit being implemented as a near-memory processor, latency for memory accesses
will be reduced compared to if a traditional compute core (CU) were to perform it. 
An example of this behavior can be
seen below, where data is able to be reliably prefetched in a timely manner due to the
dependencies clearly dictated by the program graph. The necessary input data for
Codelet \texttt{Comp1} can be prefetched from memory through Memory Codelet \texttt{LoadData1\_2048L} while Codelet \texttt{Comp0} is being executed. 
Hence, overlapping compute of \texttt{Comp0}, prefetching of data into \texttt{R2048L\_3}, followed by no DRAM memory access during execution of \texttt{Comp1}. This example is general
and can be extended to various situations. Furthermore, we envision that with the help of DMA-like hardware, the Memory Codelet 
Unit should be able to reliably feed the necessary data to the Compute Units in the
system.

\begin{center}
\begin{lstlisting}[style=customasm, label=lst:spGEMMprefetch, caption={{Sparse GEMM outer product partial prefetching example}}, firstnumber=01]
// Load data for Comp0, prefetch data for Comp1
MEMCOD LoadData0_2048L R2048L_2, R64B_6, R64B_22;
MEMCOD LoadData1_2048L R2048L_3, R64B_7, R64B_23;
// Comp0 depends on Data0 (R_2)
// Comp1 depends on Comp0 (R_1) and Data1 (R_3)
COD Comp0_2048L R2048L_1, R2048L_2;
COD Comp1_2048L R2048L_3, R2048L_1, R2048L_3;
// Store computation result
MEMCOD StoreData_2048L R2048L_3, R64B_7, R64B_23;
\end{lstlisting}
\end{center}

In programs where the access pattern within the register is known by producer and consumer, streaming
can be fully overlapped by computation in a timely fashion. To further support this
organization of execution, hardware-based FIFO queues could be added to the system
as mentioned in \cite{CodeletModelinChiplets,SidThesis}. With hardware FIFOs available,
the streaming could occur as early as allowed by the algorithm and the Codelet Model
semantics without using on-CU resources (like local memory) that might be needed by
other Codelets' execution. The FIFO queue can be used as a buffer between the memory loads performed by the
Memory Codelet Unit and the consumer Codelet performing main computation. 
Hence, traditional streaming can be 
implemented as a pipeline from a Memory Codelet to a compute Codelet via FIFO queue.

In the example below, a smart streaming based
outer product sparse GEMM application is shown. The Memory Codelets walk the 
compressed CSR and CSC formats and stream elements to the compute Codelet to perform the outer-product multiplications, producing partial matrices.
As the partial result matrices are created, we can imagine that the register that connects \texttt{spOuterMatMult} and \texttt{PartialsSum}, \texttt{R2048L\_4}, 
acts as a FIFO, such that they are streamed to the \texttt{PartialsSum} Codelet
to form the complete result matrix.

\begin{center}
\begin{lstlisting}[style=customasm, label=lst:spGEMMstreaming, caption={{Outer product sparse GEMM streaming}}, firstnumber=01]
// Stream chunks of matrix to spOuterMatMult
MEMCOD StreamCSCBlock_2048L R2048L_2, R64B_6, R64B_22;
MEMCOD StreamCSRBlock_2048L R2048L_3, R64B_7, R64B_23;
// Perform outer product mult
// Stream partial result mats. out
COD spOuterMatMult_2048L R2048L_4, R2048L_2, R2048L_3;
// Stream in partial matrices and sum
COD PartialsSum_2048L R64B_8, R2048L_4
\end{lstlisting}
\end{center}
Beyond this, Memory Codelets being used for prefetching and streaming can be extended
into performing recode operations, similar to \cite{tako, UDP2017, UAP}. A recode operation could easily be crafted by 
pipelining both prefetching/streaming and preprocessing in the Memory Codelet Unit itself.
This can be viewed in terms of batches of data: the CU will be performing main 
computation on the earliest batch of data while the Memory Codelet Unit is performing
preprocessing on the next batch and the third batch is in flight. Though these
applications of Memory Codelets are somewhat dependent on the implementation and the
specific memory hierarchy structure of the architecture in use, their concepts are
applicable in many cases and the extended Codelet Model provides a cohesive model
of program execution. An example can be seen below where the same
outer product computation is performed, but both matrices are in CSC form, so a Memory Codelet
performs recoding to change the second matrix to CSR for ease of computation. This example
maintains the same streaming format as the earlier example.

\begin{center}
\begin{lstlisting}[style=customasm, label=lst:spGEMMrecode, caption={{Outer product sparse GEMM streaming with recode operation}}, firstnumber=01]
// Fetch block of B; recode block of C into CSR format;
// stream both to CU
MEMCOD FetchCSCBlock_2048L R2048L_2, R64B_6, R64B_22;
MEMCOD ConvertCSCBlock_2048L R2048L_3, R64B_7, R64B_23;
// Do sp outer product mult; stream partial mat. out
COD spOuterMatMult_2048L R2048L_4, R2048L_2, R2048L_3;
// Sum streamed-in matrices, store result
COD PartialsSum_2048L R64B_8, R64B2
\end{lstlisting}
\end{center}

\subsection{Modification and Use of Conventional Memory Systems}
Modern memory subsystems are complex with a very large design space, and can be difficult to effectively
utilize in non conventional applications with irregular memory behavior. Though Memory Codelets 
can certainly aid effective memory use strategies in Codelet programs, their effects are largely
dependent on the memory hierarchy of the system and its protocols. Two non-exclusive paths can be
targeted to improve performance of programs based on utilization of a system's memory hierarchy: modification
of memory systems to include less-conventional components/strategies and tuning of programs to better their
use of conventional components.

\subsubsection{Tuned Use of Conventional Memory Systems}
If we assume that the program
is executing on a conventional processing system with a 3-level cache hierarchy between the processor
and DRAM, timeliness of prefetching becomes paramount due to the danger of data being evicted from the
cache early. We speculate that this issue could be mitigated thanks to Memory Codelets and the static
information present in the Codelet Graph; L1 and L2 use would be predictable based on the properties
of the Codelets. Furthermore, increasing performance of the system based on prefetching can benefit from
more targeted approaches, as in \cite{APAC-Sun}. Publications such as \cite{LPM-Sun}\cite{CAL-Sun} can
illuminate the importance of eliminating pure miss cycles in the cache hierarchy. With this information
improved scheduling mechanisms can be applied and Memory Codelets can be organized to improve bandwidth 
utilization. This is one of the major benefits of a software-programmable unit
that can perform prefetching, combined with scheduling that is memory-aware.
Though the memory hierarchy in use is detached from the Codelet Model semantics, in a conventional 
system the LLC can be thought of as Threaded Procedure memory. This indicates that ideally, before
a Codelet's firing, the data it requires should be resident on that CU's L1 or L2 cache (depending on
size). The wisdom in the citations above
could then be applied conceptually as balancing the flow of data between TP memory and CU memory,
with prediction made easier by the Codelet Program graph.

\subsubsection{Modified Memory Systems}
Despite improving effective utilization on conventional memory systems, each memory hierarchy has its
own drawbacks. In this case, consider the earlier distinction between LLC as TP memory and L1/L2 caches as
CU local memory. Codelets have well defined input and output and we expect the input data to be
local to the CU at the time of firing. This means in practice, a Codelet's ``size'' would be limited
based on the size of local memory storage. However, this can be avoided through the use of
streaming Codelets, especially with the aid of Memory Codelets and FIFO queues. This would be implemented 
best with FIFO queues in hardware, but this may require modification of the memory hierarchy and
how data is moved through it. Codelet Model semantics and the configurability provided by Memory Codelets
would allow hardware-based FIFO queues to coexist with typical memory hierarchies.

In addition to hardware FIFO queues, a hierarchy of scratchpad memory units could replace the typical
cache hierarchy, or more practically, the upper levels of it. This would allow the software to have more
control over how data is prefetched in anticipation of firing Codelets and at what granularity data
is moved between DRAM and the CUs. A scratchpad memory system would particularly aid in programs that
typically achieve low performance on conventional systems due to low use of data loaded because of
cache line size. In addition, it would avoid slowdowns in programs that stall often due to cache
thrashing and cache line invalidations. It is a possibility that both to 
maintain coherence and improve performance throughout the memory hierarchy, Memory Codelets could be inserted
into the Codelet Graph by the compiler based on analysis of the graph and the Codelets' dependencies, somewhat analogous to the strategy of \cite{prodigy} in a different PXM. 

\begin{figure}
    \centering
    \includegraphics[width=9cm]{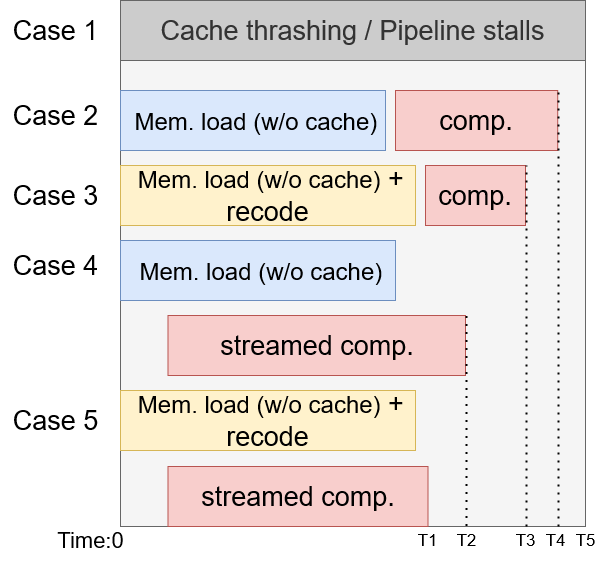}
    \caption{A representation of how Memory Codelets could be used to accelerate programs
    with poor cache behavior. Blue and yellow boxes include loading memory in a way that
    bypasses cache (e.g. loading from scratchpad memory). Yellow boxes include recoding
    operations done by the MCU such that main computation is more regular.}
    \label{fig:mem_cod_transform}
\end{figure}

Fig. \ref{fig:mem_cod_transform} shows an illustration of strategies to use Memory Codelets to improve a program
that is hindered by typical cache protocols. In this Figure T5 represents the runtime of a program
in a traditional architecture. T4 benefits from prefetching data. T3 is lowered by
prefetching data and recoding it, such that the computation itself is faster. T2 uses
streaming from a Memory Codelet to the computation Codelet via FIFO. Finally T1
combines all the approaches: Prefetching, recoding, and streaming.

\subsubsection{Memory Codelets and Coherency}
Because Memory Codelets can allow for fine-grained control over data movement throughout the 
memory hierarchy of the system, they can help the Codelet Model to fulfill the role of 
providing coherency to the user at a software level. 
Most conventional systems with multiple compute units on a
chip (such as multiple cores in a multiprocessor) have mechanisms to provide memory coherency
throughout the memory hierarchy and between the compute units. Coherency at the hardware
level allows the software to have a flat, one dimensional view of the memory without having
to manage multiple copies of data. 
While this is certainly useful for programmers developing multithreaded applications, the 
hardware
mechanisms that provide the coherency can use up precious chip real estate and the cost can
scale poorly as the number of compute units and physical memory units increase on a chip. 
Beyond this, enforcing the coherency model through cache line invalidation can reduce effective
use of memory resources.
The need for hardware coherency can be thought of as a crutch for programmers using the
threading model for parallel programming; in other words, the threading model of computation
is so ill-defined and prone to non-determinism that programmers would have great difficulty
developing on a multithreaded system without hardware coherency \cite{LeeProblemWithThreads}. However, the situation
depends entirely on the semantics of the Program Execution Model employed in the system. 

In the threading model of parallel programming, various threads can concurrently execute and access
the entire memory space of the program. This places the entire burden of synchronization and
avoiding data races on the programmer's shoulders, which is notoriously difficult
\cite{LeeProblemWithThreads}. 
Furthermore, on
even the most modestly heterogeneous systems, coherency generally becomes the responsibility of the
programmer through explicit data movements between host and device.

In the Codelet Model PXM, computation is broken down into pieces of computation that are
partially ordered by data dependencies between them as expressed in the Codelet Graph. 
Through this mechanism, data races are avoided
and the necessary synchronization is achieved. Moreover, because the program is broken down into
Codelets and their specific input and output data, two or more Codelets that access the same data (with
at least one of the accesses being a write/store) will only execute when no data dependency has been
declared between them, which in a correctly written program indicates the program is intended to have
non-deterministic behavior in that program section. In other words, Codelets are partially ordered
based on their data dependencies. 

Readers may question the difference in difficulty
between writing synchronization into a multithreaded program and correctly writing the dependencies
in a Codelet Model program; the main difference lies in the finite behavior of Codelets and the data
they access. Synchronization and orchestration of threads that can access any data in the address
space is unwieldy and unmanageable, whereas finer-grained Codelets with well defined data access
make the process more straightforward. As systems trend towards extreme heterogeneity, hardware coherency
may not be feasible. The CXL 3.0 specification includes coherency mechanisms \cite{CXL} but no
implementations have been released yet. While this is generally considered manageable in a CPU-GPU
only system, with various accelerators or specialized architectures, each possibly
having their own memory hierarchy it rapidly becomes less manageable. The burden is eased with
a well defined PXM and Memory Codelets to provide clear functionality. This issue is discussed
further with respect to chiplet-based systems in \cite{CodeletModelinChiplets}.

%% file: tex/conclusion.tex
\section{Conclusion}
In this paper we introduce the concept of Memory Codelet into the Codelet Model. We define an extended abstract machine with a Memory Codelet Unit that supports coordination of prefetching, streaming and memory scheduling operations, as well as data recoding operations. We demonstrate through the use of the Sequential Codelet Model and how Memory Codelets can be used in the context of an application. Future work includes implementation and testing of this work on heterogeneous
architectures.